\documentclass[gmd, manuscript]{copernicus}

\usepackage{graphicx}
\allowdisplaybreaks
\nolinenumbers

\begin{document}

\title{FORTVSH v1.0.0: A Fortran Library For Vector Spherical Harmonics Computations}

%\Author[1]{Justin G.}{Elfritz\thanks{Current address: IBM, Poughkeepsie, NY, USA}}
\Author[1\thanks{Present Address: IBM, Pittsburgh, PA, USA}][justin.elfritz@gmail.com]{Justin G.}{Elfritz}
% Correspondence author definition
%\assignauthor{1}{justin.elfritz@gmail.com} 

% Affiliations
\affil[1]{Independent Researcher, Pittsburgh, PA, USA}

%\correspondence{Justin G. Elfritz (justin.elfritz@gmail.com)}
%% The [] brackets identify the author with the corresponding affiliation. 1, 2, 3, etc. should be inserted.

%% If an author is deceased, please add \deceased[$Deceased date if applicable$]{$Author number$} (e.g. \deceased[13 November 2015]{2}) at the end of the affiliations. The author number depends on the placement of the author in the author list, e.g. the third author has number 3.

%% If authors contributed equally, please add \equalcontrib{$Author numbers$} (e.g. \equalcontrib{1,3}) at the end of the affiliations. The author number depends on the placement of the author in the author list, e.g. the third author has number 3.
\runningtitle{FORTVSH}

\runningauthor{J.G.\,Elfritz}

\received{}
\pubdiscuss{} %% only important for two-stage journals
\revised{}
\accepted{}
\published{}

%% These dates will be inserted by Copernicus Publications during the typesetting process.

\firstpage{1}

\maketitle

\begin{abstract}
The vector spherical harmonics (VSH) are crucial ingredients for representation of arbitrary vector fields in 
curvilinear coordinate systems throughout mathematical physics. Performant, efficient computations of the VSH are essential 
to high-performance computing applications that utilize spectral analysis, including magnetohydrodynamics (MHD) simulations 
for geophysics and astrophysics. The free open-source FORTVSH library provides a catalog of numerical routines for evaluating various VSH forms and related mathematical functions. This manuscript establishes the theoretical basis, addresses the computational need for efficient VSH solvers, and presents a comprehensive overview of the FORTVSH routines.
\end{abstract}

%\copyrightstatement{TEXT} %% This section is optional and can be used for copyright transfers.

\introduction  %% \introduction[modified heading if necessary]
The tensor spherical harmonics (TSH) are irreducible tensor products of scalar spherical harmonics $Y_L^M$ 
and basis spin functions $\chi_{S \sigma}$ \citep{QTAM}. An arbitrary TSH of rank $S$ written as $\mathbf{Y}_{J M}^{L S}$ may be expressed as 
a sum over eigenstates and state amplitudes as
\begin{equation}\label{eq:TSHexpansion}
  \mathbf{Y}_{J M}^{L S} = \sum_{m \sigma} C_{L m S \sigma}^{J M} Y_{L}^{M}\chi_{S {\sigma}}
\end{equation}
  
\noindent where $C_{a \alpha b \beta}^{c \gamma}$ is a Clebsch-Gordan coefficient, and $S$ takes non-negative integer or half-integer values. 
This notation emphasizes the coupling between orbital angular momentum $\vec{L}$ and spin angular momentum $\vec{S}$ from which we 
determine the total angular momentum $\vec{J} = \vec{L}+\vec{S}$. In the usual spherical polar coordinate system, we know $Y_L^M$ to be 
eigenfunctions of $L^2$ and $L_z$, and $\chi_{S {\sigma}}$ are eigenfunctions of $S^2$ and $S_z$ \citep{Edmonds1960}. The particular case of interest here 
is for tensor rank $S=1$, in which case the expansion Eq.\,(\ref{eq:TSHexpansion}) becomes
\begin{equation}\label{eq:VSHexpansion}
    \mathbf{Y}_{J M}^{L} = \sum_{m \sigma} C_{L m 1 \sigma}^{J M} Y_{L}^{m}\hat{e}_{\sigma}
\end{equation}

\noindent where $\hat{e}_{\sigma}$ are unit covariant spherical basis vectors (Spin-1 eigenstates). Selection rules determine the 
non-zero Clebsch-Gordan coefficients, and in this case $L=J,J\pm 1$ and $\sigma=0,\pm 1$. The $\mathbf{Y}_{J M}^{L}$ in 
Eq.\,(\ref{eq:VSHexpansion}) represent the vector spherical harmonics (VSH), which are a limiting case of the general TSH in 
Eq\,(\ref{eq:TSHexpansion}). Throughout this text we 
adopt the following normalization for $Y_\ell^m$:
\begin{equation}
  Y_{\ell}^m = \sqrt\frac{2\ell+1}{4\pi}\sqrt{\frac{\left(\ell-m\right)!}{\left(\ell+m\right)!}}P_\ell^m(\cos\theta)e^{im\phi}
\end{equation}

\noindent where $P_\ell^m(\cos\theta)$ is the associated Legendre polynomial.

One broad computational objective of this work is to facilitate time-resolved numerical simulations of magnetohydrodynamic (MHD) systems in spherical 
geometry. The VSH in Eq.\,(\ref{eq:VSHexpansion}) constitute a complete orthonormal basis, and therefore enable
spectral expansion of 3D MHD vector fields with arbitrary angular configuration. The focus of this paper is to describe and validate the
computation of special forms of the VSH that are useful for representation of MHD magnetic fields decomposed into poloidal and toroidal components. Designing MHD simulation infrastructure is out of scope of this paper, but future simulation codes may be built with FORTVSH as a key architectural element. 

Several open-source Fortran packages are available for working with spherical harmonics expansions and transformations, most notably SHTOOLS \citep{SHTOOLS} and SPHEREPACK \citep{spherepack3}, however neither directly supports poloidal-toroidal magnetic field representation (see Section\,\ref{s:discussion}). FORTVSH provides a lightweight, physics-first approach that preserves algebraic conventions from the applied physics literature and reduces compute overhead in the vector potential formulation. The low-level implementation of FORTVSH routines provides limited abstraction, making FORTVSH immediately suitable for numerical experiments in many system configurations, including but not limited to 3D coupled spherical shells (\textit{e.g.,} core-crust interactions in neutron stars), 2D or 3D stellar or planetary boundary matching studies, modeling 2D systems with azimuthal symmetry, or relaxation algorithms. 

To achieve the desired poloidal-toroidal decomposition, the $\mathbf{Y}_{JM}^{(\lambda)}$ form of the VSH are leveraged 
following \cite{AkheizerBook} and \cite{QTAM}. 
These $\mathbf{Y}_{JM}^{(\lambda)}$ are not eigenfunctions of $L^2$ 
but they form a complete orthonormal basis and are linear combinations of the $\mathbf{Y}_{JM}^{L}$. Table \ref{t:vshsummary} 
provides a summary of the relationship between the scalar spherical harmonics $Y_J^M$, the alternate VSH $\mathbf{Y}_{JM}^{(\lambda)}$, 
the $L^2$ eigenfunction VSH $\mathbf{Y}_{JM}^{L}$, and the VSH notation presented by \cite{Barrera1985}. The shorthand 
$\sqrt{\Lambda_J}=\sqrt{J(J+1)}$ also appears in Table \ref{t:vshsummary} and subsequent sections, as well as $R_J = r/\sqrt{J(J+1)}$. 
\begin{table}[h]
    \begin{center}
        \begin{tabular}{| c | c | c | c |}
            \hline
            \textbf{Scalar} & \textbf{VSH} & $\boldsymbol{\it L^2}$ \textbf{VSH} & \textbf{Barrera et al.} \\
            \hline
            $\hat{r}Y_J^M$ & $\mathbf{Y}_{J M}^{(-1)}$ & $\frac{1}{\sqrt{2J+1}}\left[\sqrt{J}\mathbf{Y}_{J M}^{J-1} - \sqrt{J+1}\mathbf{Y}_{J M}^{J+1}\right]$ & $\mathbf{Y}_{J M}^{}$ \\
            $-\frac{i}{\sqrt{\Lambda_J}}\hat{r}\times\vec{\nabla}_{\omega} Y_J^M$ & $\mathbf{Y}_{J M}^{(0)}$ & $\mathbf{Y}_{J M}^{J}$ & $-\frac{i}{r\sqrt{\Lambda_J}}\mathbf{\Phi}_{J M}$ \\
            $\frac{1}{\sqrt{\Lambda_J}}\vec{\nabla}_{\omega} Y_J^M$ & $\mathbf{Y}_{J M}^{(+1)}$ & $\frac{1}{\sqrt{2J+1}}\left[\sqrt{J+1}\mathbf{Y}_{J M}^{J-1} + \sqrt{J}\mathbf{Y}_{J M}^{J+1}\right]$ & $\frac{1}{r\sqrt{\Lambda_J}}\mathbf{\Psi}_{J M}$ \\
            \hline
        \end{tabular}
        \caption{Summary of notations and forms of vector spherical harmonics (VSH). Each row contains four equivalent entries, 
        from left to right: vector calculus formula with scalar spherical harmonics, 
        the alternate VSH notation used in this paper, the VSH formulation using eigenfunctions of $L^2$, 
        and the notation of \cite{Barrera1985}.}
        \label{t:vshsummary}
    \end{center}
\end{table}

Both forms of the VSH from Table \ref{t:vshsummary} permit series expansions of angular functions $F(\theta,\phi)$ due to completeness and orthonormality: 
\begin{align}
F(\theta,\phi) = \sum_{J L M} A_{J M}^{L} \mathbf{Y}_{J M}^{L}(\theta,\phi) \\
F(\theta,\phi) = \sum_{J \lambda M} A_{J M}^{\left(\lambda\right)} \mathbf{Y}_{J M}^{(\lambda)}(\theta,\phi)
\end{align}

\noindent where the expansion coefficients are 
\begin{align}
A_{J M}^{L} = \int_{4\pi}\mathrm{d}\Omega\,F(\theta,\phi) \mathbf{Y}_{J M}^{L*}(\theta,\phi) \label{e:vshcoeff1}\\
A_{J M}^{\left(\lambda\right)} = \int_{4\pi}\mathrm{d}\Omega\,F(\theta,\phi) \mathbf{Y}_{J M}^{(\lambda *)}(\theta,\phi)\label{e:vshcoeff2}
\end{align}

Solving the magnetic induction equation $\partial_t\vec{B}=-c\vec{\nabla}\times\vec{E}$ is the key challenge in most MHD numerical simulations. The detailed form of the induction equation informs the algorithm design and implementation, and will reflect the specific physics regime being studied. This computation problem in the spherical 3D MHD context may be simplified by decomposing the magnetic vector field $\vec{B}$ into two simultaneous 2D problems representing the poloidal and toroidal field components \citep{KrauseRadlerBook}. This is achieved by utilizing the magnetic vector potential $\vec{A}$, which satisfies $\vec{B}=\vec{\nabla}\times\vec{A}$, and exploiting the purely rotational character of the curl operator:
\begin{align}
\partial_t\vec{B}_\mathrm{tor} = & \partial_t\left[\vec{\nabla}\times\left(\vec{r}\Psi\left(\vec{r},t\right)\right)\right] = \partial_t\left(-\vec{r}\times\vec{\nabla}\Psi\left(\vec{r},t\right)\right) = -\vec{r}\times\vec{\nabla}\left(\partial_t\Psi\left(\vec{r},t\right)\right) = -c\vec{\nabla}\times\vec{E}_\mathrm{pol} \\
\partial_t\vec{B}_\mathrm{pol} = & \vec{\nabla}\times\partial_t\vec{A}_\mathrm{tor} = \vec{\nabla}\times\partial_t\left[\vec{\nabla}\times\left(\vec{r}\Phi\left(\vec{r},t\right)\right)\right] = -\vec{\nabla}\times\left[\vec{r}\times\vec{\nabla}\left(\partial_t\Phi\left(\vec{r},t\right)\right)\right] = -c\vec{\nabla}\times\vec{E}_\mathrm{tor}
\end{align}

\noindent Here $\Phi\left(\vec{r},t\right)$ and $\Psi\left(\vec{r},t\right)$ are the time-dependent poloidal and toroidal scalar potentials respectively, not to be confused with any scalar potentials used in magnetostatics. $\Phi\left(\vec{r},t\right)$ and $\Psi\left(\vec{r},t\right)$ are used here to generate toroidal components of $\vec{B}$ and $\vec{A}$ for dimensionality reduction, and the eigenmode amplitudes of these two potentials will be iteratively updated in the MHD time advance algorithm. Most generally, 
\begin{align}
\Phi\left(\vec{r},t\right) = & \frac{1}{r}\sum_{\ell,m}\Phi_{\ell,m}\left(r,t\right)Y_{\ell}^m\left(\theta,\phi\right) \\
\Psi\left(\vec{r},t\right) = & \frac{1}{r}\sum_{\ell,m}\Psi_{\ell,m}\left(r,t\right)Y_{\ell}^m\left(\theta,\phi\right)
\end{align}
\noindent and thus
\begin{align}
\vec{A}_\mathrm{tor}\left(\vec{r},t\right) = -\frac{i}{r}\sum_{\ell,m}\sqrt{\ell\left(\ell+1\right)}\Phi_{\ell m}\left(r,t\right)\vec{\mathrm{Y}}_{\ell m}^{\left(0\right)}\\
\vec{B}_\mathrm{tor}\left(\vec{r},t\right) = -\frac{i}{r}\sum_{\ell,m}\sqrt{\ell\left(\ell+1\right)}\Psi_{\ell m}\left(r,t\right)\vec{\mathrm{Y}}_{\ell m}^{\left(0\right)}
\end{align}

Developing a self-consistent numerical MHD formulation using this approach requires a clear understanding of how differential operators transform VSH modes since VSH spectra will propagate through the induction equation, self-couple, and ultimately power the system's time evolution. The specifics require an analysis of the MHD regime and associated expansion of the induction electric field. Algorithm design and numerical simulation of various MHD limits (\textit{e.g.}, simple diffusion, nonlinear diffusion, Hall drift) are beyond the scope of this paper, and are left for future development work outside of FORTVSH.

The contents of this manuscript are organized as follows. The analytic foundation of the FORTVSH codebase and the connection to other standard representations are found in Section\,\ref{s:formulation}. Useful mathematical identities relating the various VSH forms are summarized in Section\,\ref{s:vshidentities}. Section\,\ref{s:cgidentities} contains identities related to Clebsch-Gordan coefficients, Wigner symbols, as well as special cases of useful relationships. Section\,\ref{s:validationapplications} provides visibility into FORTVSH testing, performance and convergence benchmarking, and overall numerical validation. Subsection\,\ref{s:validation} presents low-level codebase validation and reports for numerical tolerances of core module functionality, and Section\,\ref{s:applications} contains four FORTVSH demonstration use cases. Section\,\ref{s:discussion} contains an abbreviated discussion of the 
physical systems subject to future numerical modeling efforts, as well as the assumptions, conditions, and parameterizations necessary to permit simulation of those systems, and finally a capabilities comparison with other widely-used spherical harmonics packages. Appendix \ref{a:numericalmodule} provides a detailed technical description of the FORTVSH routines.

\section{Analytic Formulation}\label{s:formulation}

\subsection{Identities involving Vector Spherical Harmonics}\label{s:vshidentities}

\noindent Equations (\ref{e:firstcurl}) - (\ref{e:finalcurl}) show the curl of VSH modes coupled to separable radial functions $f(r)$:
\begin{align}
\vec{\nabla}\times\left(f(r)\mathbf{Y}_{JM}^{(-1)}\right) = & -\frac{i}{R_J}f(r)\mathbf{Y}_{JM}^{(0)} \label{e:firstcurl}\\
\vec{\nabla}\times\left(f(r)\mathbf{Y}_{JM}^{(+1)}\right) = & \;\frac{i}{R_J}\frac{d}{dr}\left(R_Jf(r)\right)\mathbf{Y}_{JM}^{(0)} \\
\vec{\nabla}\times\left(f(r)\mathbf{Y}_{JM}^{(0)}\right) = & \;\frac{i}{R_J}\left[f(r)\mathbf{Y}_{JM}^{(-1)}+\frac{d}{dr}\left(R_Jf(r)\right)\mathbf{Y}_{JM}^{(+1)}\right]
\end{align}
\begin{align}
    \vec{\nabla}\times\left(f(r)\mathbf{Y}_{JM}^{J-1}\right) = & \sqrt{\frac{J+1}{2J+1}}\frac{i}{R_J^{1-J}}\frac{d}{dr}\left(R_J^{1-J}f(r)\right)\mathbf{Y}_{JM}^{J}\\
    \vec{\nabla}\times\left(f(r)\mathbf{Y}_{JM}^{J+1}\right) = & \sqrt{\frac{J}{2J+1}}\frac{i}{R_J^{J+2}}\frac{d}{dr}\left(R_J^{J+2}f(r)\right)\mathbf{Y}_{JM}^{J}\\
    \vec{\nabla}\times\left(f(r)\mathbf{Y}_{JM}^{J}\right) = & \frac{i}{\sqrt{2J+1}}\left[\frac{\sqrt{J+1}}{R^{J+1}}\frac{d}{dr}\left(R^{J+1}f(r)\right)\mathbf{Y}_{JM}^{J-1} + \frac{\sqrt{J}}{R^{-J}}\frac{d}{dr}\left(R^{-J}f(r)\right)\mathbf{Y}_{JM}^{J+1}\right]\label{e:finalcurl}
\end{align}

\noindent The Clebsch-Gordan series is 
\begin{align}\label{e:clebschgordanseries}
    \mathbf{Y}_{J_1 M_1}^{L_1}\cdot\mathbf{Y}_{J_2 M_2}^{L_2} = & \sum_{L'} (-1)^{J2+L1+{L'}}\sqrt{\frac{\left(2J_1+1\right)\left(2J_2+1\right)\left(2L_1+1\right)\left(2L_2+1\right)}{4\pi\left(2{L'}+1\right)}}\times\nonumber\\
    & \begin{Bmatrix} L_1 & L_2 & {L'} \\ J_2 & J_1 & 1 \end{Bmatrix}C_{L_1 0 L_2 0}^{{L'} 0}C_{J_1 M_1 J_2 M_2}^{{L'} M_1+M_2}Y_{L'}^{M_1+M_2}, 
\end{align}

\noindent which is used to compute the inner products of VSH modes. The term in curly braces represents the Wigner 6-j symbol.  

The complete set of VSH inner products is 
compiled here in the condensed $I_{k'l'kl}^{nm}, J_{k'l'kl}^{nm}$ notation\footnote{Disambiguation note: 
The author preserves the original $J$ notation of the referenced authors, not to be confused with the total
angular momentum parameter $J$ indexing the VSH $\mathbf{Y}_{J M}^L$ elsewhere in this manuscript.} of 
\cite{GW1} (see also Section\,\ref{s:cgidentities}). Inner products of the $L^2$ VSH modes are
\begin{align}
    \mathbf{Y}_{kl}^{k}\cdot\mathbf{Y}_{nm}^{n} = & -\sum_{L'}\frac{\Lambda_k + \Lambda_n - \Lambda_{L'}}{2\sqrt{\Lambda_k\Lambda_n}}I_{k l n m}^{{L'} l+m}Y_{L'}^{l+m} \label{e:firstdotprod1}\\
    \mathbf{Y}_{kl}^{k}\cdot\mathbf{Y}_{nm}^{n-1} = & \frac{-i}{\sqrt{\Lambda_k\Lambda_n}}\sqrt{\frac{n+1}{2n+1}}\sum_{L'} J_{k l n m}^{{L'} l+m}Y_{L'}^{l+m} \\
    \mathbf{Y}_{kl}^{k}\cdot\mathbf{Y}_{nm}^{n+1} = & \frac{-i}{\sqrt{\Lambda_k\Lambda_n}}\sqrt{\frac{n}{2n+1}}\sum_{L'} J_{k l n m}^{{L'} l+m}Y_{L'}^{l+m} \\
    \mathbf{Y}_{kl}^{k-1}\cdot\mathbf{Y}_{nm}^{n-1} = & \sqrt{\frac{kn}{(2k+1)(2n+1)}}\sum_{L'}\left[\frac{\Lambda_k+\Lambda_n-\Lambda_{L'}}{2}+1\right]I_{k l n m}^{{L'} l+m}Y_{L'}^{l+m} \\
    \mathbf{Y}_{kl}^{k-1}\cdot\mathbf{Y}_{nm}^{n+1} = & \sqrt{\frac{k(n+1)}{(2k+1)(2n+1)}} \sum_{L'}\left[\frac{\Lambda_k+\Lambda_n-\Lambda_{L'}}{2k(n+1)}-1\right]I_{k l n m}^{{L'} l+m}Y_{L'}^{l+m}\\
    \mathbf{Y}_{kl}^{k+1}\cdot\mathbf{Y}_{nm}^{n+1} = & \sqrt{\frac{(k+1)(n+1)}{(2k+1)(2n+1)}} \sum_{L'}\left[\frac{\Lambda_k+\Lambda_n-\Lambda_{L'}}{2(k+1)(n+1)}+1\right]I_{k l n m}^{{L'} l+m}Y_{L'}^{l+m}\label{e:finaldotprod1}
\end{align}

\noindent and inner products of the alternate VSH modes are 
\begin{align}    
    \mathbf{Y}_{kl}^{(-1)}\cdot\mathbf{Y}_{nm}^{(-1)} = & \sum_{L'} I_{k l n m}^{{L'} l+m} Y_{L'}^{l+m}\label{e:firstdotprod2} \\
    \mathbf{Y}_{kl}^{(-1)}\cdot\mathbf{Y}_{nm}^{(0)} = & \quad 0 \\
    \mathbf{Y}_{kl}^{(-1)}\cdot\mathbf{Y}_{nm}^{(+1)} = & \quad 0 \\
    \mathbf{Y}_{kl}^{(0)}\cdot\mathbf{Y}_{nm}^{(0)} = & -\sum_{L'}\frac{\Lambda_k + \Lambda_n - \Lambda_{L'}}{2\sqrt{\Lambda_k\Lambda_n}}I_{k l n m}^{{L'} l+m}Y_{L'}^{l+m} \\
    \mathbf{Y}_{kl}^{(+1)}\cdot\mathbf{Y}_{nm}^{(0)} = & \frac{-i}{\sqrt{\Lambda_k\Lambda_n}}\sum_{L'} J_{n m k l}^{{L'} l+m} Y_{L'}^{l+m}\\
    \mathbf{Y}_{kl}^{(+1)}\cdot\mathbf{Y}_{nm}^{(+1)} = & \sum_{L'}\frac{\Lambda_k + \Lambda_n - \Lambda_{L'}}{2\sqrt{\Lambda_k\Lambda_n}}I_{k l n m}^{{L'} l+m}Y_{L'}^{l+m} \label{e:finaldotprod2}
\end{align}

%%%%%%%%%%%%%%%%%%%%%%%%%%%%%%%%%5

\noindent The complex conjugates of VSH modes are
\begin{align}
    \mathbf{Y}_{J M}^{L *} = & (-1)^{J + L + M + 1}\mathbf{Y}_{J -M}^{L} \\
    \mathbf{Y}_{J M}^{(\lambda)*} = & (-1)^{\lambda + M + 1}\mathbf{Y}_{J -M}^{(\lambda)}
\end{align}

\noindent which gives a special case of the Clebsch-Gordan series
\begin{align}\label{e:conjclebschgordanseries}
    \mathbf{Y}_{J_1 M_1}^{L_1}\cdot\mathbf{Y}_{J_2 M_2}^{L_2 *} = & \sum_{L'} (-1)^{L_1+L_2-M_1}\sqrt{\frac{\left(2J_1+1\right)\left(2J_2+1\right)\left(2L_1+1\right)\left(2L_2+1\right)}{4\pi\left(2{L'}+1\right)}}\times\nonumber\\
    & \begin{Bmatrix} L_1 & L_2 & {L'} \\ J_2 & J_1 & 1 \end{Bmatrix}C_{L_1 0 L_2 0}^{{L'} 0}C_{J_1 M_1 J_2 -M_2}^{{L'} M_1-M_2}Y_{L'}^{M_2-M_1 *}
\end{align}

\noindent In combination with the results in Eqs.\,(\ref{e:firstdotprod1}--\ref{e:finaldotprod1}), the 
Hermitian inner products of the $L^2$ VSH are
\begin{align}
  \mathbf{Y}_{kl}^{k}\cdot\mathbf{Y}_{nm}^{n *} = & \sum_{L'}\frac{\Lambda_k+\Lambda_n-\Lambda_{L'}}{2\sqrt{\Lambda_k\Lambda_n}}I_{k l {L'} m-l}^{n m}Y_{L'}^{m-l *} \\
  \mathbf{Y}_{kl}^{k}\cdot\mathbf{Y}_{nm}^{n-1 *} = & \frac{i}{\sqrt{\Lambda_k \Lambda_n}}\sqrt{\frac{n+1}{2n+1}}\sum_{L'} J_{k l {L'} m-l}^{n m} Y_{L'}^{m-l *} \\
  \mathbf{Y}_{kl}^{k}\cdot\mathbf{Y}_{nm}^{n+1 *} = & \frac{i}{\sqrt{\Lambda_k \Lambda_n}}\sqrt{\frac{n}{2n+1}}\sum_{L'} J_{k l {L'} m-l}^{n m} Y_{L'}^{m-l *} \\
  \mathbf{Y}_{kl}^{k-1}\cdot\mathbf{Y}_{nm}^{n *} = & \frac{i}{\sqrt{\Lambda_k \Lambda_n}}\sqrt{\frac{k+1}{2k+1}}\sum_{L'} J_{k l {L'} m-l}^{n m} Y_{L'}^{m-l *} \\
  \mathbf{Y}_{kl}^{k-1}\cdot\mathbf{Y}_{nm}^{n-1 *} = & \sqrt{\frac{kn}{(2k+1)(2n+1)}}\sum_{L'}\left[\frac{\Lambda_k+\Lambda_n-\Lambda_{L'}}{2}+1\right]I_{k l {L'} m-l}^{n m}Y_{L'}^{m-l *} \\
  \mathbf{Y}_{kl}^{k-1}\cdot\mathbf{Y}_{nm}^{n+1 *} = & \sqrt{\frac{k(n+1)}{(2k+1)(2n+1)}} \sum_{L'}\left[\frac{\Lambda_k+\Lambda_n-\Lambda_{L'}}{2k(n+1)}-1\right]I_{k l {L'} m-l}^{n m}Y_{L'}^{m-l *} \\
  \mathbf{Y}_{kl}^{k+1}\cdot\mathbf{Y}_{nm}^{n *} = & \frac{i}{\sqrt{\Lambda_k \Lambda_n}}\sqrt{\frac{k}{2k+1}}\sum_{L'} J_{k l {L'} m-l}^{n m} Y_{L'}^{m-l *}\\
  \mathbf{Y}_{kl}^{k+1}\cdot\mathbf{Y}_{nm}^{n-1 *} = & \sqrt{\frac{n(k+1)}{(2k+1)(2n+1)}} \sum_{L'}\left[\frac{\Lambda_k+\Lambda_n-\Lambda_{L'}}{2n(k+1)}-1\right]I_{k l {L'} m-l}^{n m}Y_{L'}^{m-l *} \\
  \mathbf{Y}_{kl}^{k+1}\cdot\mathbf{Y}_{nm}^{n+1 *} = & \sqrt{\frac{(k+1)(n+1)}{(2k+1)(2n+1)}} \sum_{L'}\left[\frac{\Lambda_k+\Lambda_n-\Lambda_{L'}}{2(k+1)(n+1)}+1\right]I_{k l {L'} m-l}^{n m}Y_{L'}^{m-l *} \\
\end{align}

\noindent and using the results Eqs.\,(\ref{e:firstdotprod2}--\ref{e:finaldotprod2}) the Hermitian inner products of the alternate VSH are written 
\begin{align}
  \mathbf{Y}_{kl}^{(-1)}\cdot\mathbf{Y}_{nm}^{(-1)*} = & \sum_{L'} I_{k l {L'} m-l}^{n m}Y_{L'}^{m-l*} \\
  \mathbf{Y}_{kl}^{(-1)}\cdot\mathbf{Y}_{nm}^{(0)*} = & \quad 0 \\
  \mathbf{Y}_{kl}^{(-1)}\cdot\mathbf{Y}_{nm}^{(+1)*} = & \quad 0 \\
  \mathbf{Y}_{kl}^{(0)}\cdot\mathbf{Y}_{nm}^{(-1)*} = & \quad 0 \\
  \mathbf{Y}_{kl}^{(0)}\cdot\mathbf{Y}_{nm}^{(0)*} = & \sum_{L'}\frac{\Lambda_k+\Lambda_n-\Lambda_{L'}}{2\sqrt{\Lambda_k\Lambda_n}}I_{k l {L'} m-l}^{n m}Y_{L'}^{m-l *}\\
  \mathbf{Y}_{nm}^{(0)}\cdot\mathbf{Y}_{kl}^{(+1)*} = & \frac{i}{\sqrt{\Lambda_k\Lambda_n}}\sum_{L'} J_{n m {L'} l-m}^{k l}Y_{L'}^{l-m *} \\
  \mathbf{Y}_{kl}^{(+1)}\cdot\mathbf{Y}_{nm}^{(-1)*} = & \quad 0 \\
  \mathbf{Y}_{kl}^{(+1)}\cdot\mathbf{Y}_{nm}^{(0)*} = & \frac{i}{\sqrt{\Lambda_k\Lambda_n}}\sum_{L'} J_{k l {L'} m-l}^{n m}Y_{L'}^{m-l *} \\
  \mathbf{Y}_{kl}^{(+1)}\cdot\mathbf{Y}_{nm}^{(+1)*} = & \sum_{L'}\frac{\Lambda_k+\Lambda_n-\Lambda_{L'}}{2\sqrt{\Lambda_k\Lambda_n}}I_{k l {L'} m-l}^{n m}Y_{L'}^{m-l *}
\end{align}

\subsection{Notation and special identities involving Clebsch-Gordan and Wigner's $3-J$ and $6-J$ coefficients}\label{s:cgidentities}

The factors $I_{k'\,l'\,k\,l}^{n\,m}$ and $J_{k' l' k l}^{n m}$ appearing throughout Section\,\ref{s:vshidentities} were introduced by \cite{GW1} as a compact
shorthand for encoding generation and amplification amplitudes associated with discrete two-mode interactions in the Hall MHD context. Spectral cascade can induce or affect
axisymmetric components ($I_{k'\,l'\,k\,l}^{n\,m}$) and non-axisymmetric components ($J_{k'\,l'\,k\,l}^{n\,m}$) based on symmetry considerations and initial conditions,
which constrain system evolution via selection rules. 

The identities in Eq.\,(\ref{e:GWIcoeff}) through (\ref{e:finalIdentity}) describing $(k,l)\leftrightarrow(k',l')$ interchange, Wigner's $6-J$ symbols of special interest, 
and Clebsch-Gordan coefficients of special interest are important for reconstructing the fundamental VSH relations presented in Section\,\ref{s:vshidentities} \citep{QTAM}
and standardizing the theoretical and computational framework.

\begin{align}
I_{k'\,l'\,k\,l}^{n\,m} = & \sqrt{\frac{(2k'+1)(2k+1)}{4\pi(2n+1)}}C_{k'\,0\,k\,0}^{n\,0}C_{k'\,l'\,k\,l}^{n m} \label{e:GWIcoeff}\\
I_{k'\,l'\,k\,l}^{n\,m} = & (-1)^{l'}I_{k'l' n -m}^{k -l} = I_{k\,l\,k'\,l'}^{n\,m}\\
J_{k' l' k l}^{n m} = & -\frac{i}{2}\sqrt{\frac{(2k'+1)(2k+1)}{4\pi(2n+1)}}\sqrt{(k'+k+n+2)(k+n-k')(k'+k-n+1)(k'-k+n+1)}\nonumber\\
 & \times C_{k'+1 0 k 0}^{n 0} C_{k' l' k l}^{n m} \label{e:GWJcoeff}\\
J_{k'\,l'\,k\,l}^{n\,m} = & -(-1)^{l'}J_{k'\,l'\,n\,-m}^{k\,-l} = -J_{k\,l\,k'\,l'}^{n\,m}
\end{align}    

\begin{align}
  \begin{Bmatrix} a & b & a+b \\ d & e & f\end{Bmatrix} = & (-1)^{a+b+d+e} \sqrt{\frac{(2a)!(2b)!(a+b+d+e+1)!(a+b-d+e)!(a+b+d-e)!}{(2a+2b+1)!(-a-b+d+e)!(a+e-f)!(a-e+f)!(a+e+f+1)!}}\times\nonumber\\
    & \;\;\;\;\;\;\times\sqrt{\frac{(-a+e+f)!(-b+d+f)!}{(b+d-f)!(b-d+f)!(b+d+f+1)!}}
    \\
  \begin{Bmatrix} a & a & 1 \\ b & b & f\end{Bmatrix} = & 2(-1)^{a+b+f+1}\sqrt{\frac{(2a-1)!(2b-1)!}{(2a+2)!(2b+2)!}}(a(a+1)+b(b+1)-f(f+1))\\
  C_{k 0 p-1 0}^{n 0} = & -C_{k 0 p+1 0}^{n 0}\sqrt{\frac{(k+n+p+2)(k+p-n+1)(k+n-p)(-k+n+p+1)}{(k+n+p+1)(k+p-n)(k+n-p+1)(-k+p+n)}}\\
  C_{k+1 0 n 0}^{L 0} = & C_{k 0 n+1 0}^{L 0}\sqrt{\frac{(k-n+L)(-k+n+L+1)}{(k-n+L+1)(-k+n+L)}}\label{e:finalIdentity}
\end{align}

\section{Numerical Validation and Applications}\label{s:validationapplications}

This section details testing, validation, and benchmarking of FORTVSH library functions and subroutines. See Appendix\,\ref{a:numericalmodule} for an overview of the numerical approach, parameterizations, and available outputs.

Timing data in this section were measured on an AMD Ryzen 7 5700G (8 cores / 16 threads) running Ubuntu 22.04.5 LTS, compiled with GNU Fortran 11.4.0 at optimization level \texttt{-O3}, applied within the Release build's \texttt{CMakeLists.txt}. Each measurement uses single-threaded \texttt{SYSTEM\_CLOCK} timing with the full grid sweep at a given $\ell_{\max}$ repeated until elapsed time exceeds 0.2\,s then reported as time per evaluation. Build, test, and execution of the full package, including the benchmark suite, were additionally verified in a container running Ubuntu 20.04 LTS with GNU Fortran 9.4.0 and CMake 3.16.3, the minimum toolchain this package requires, confirming the results are not an artifact of a specific compiler version. The timings reported in this section reflect the primary development machine above.

\subsection{Numerical Validation}\label{s:validation}

The FORTVSH test suite (\texttt{tests.f90}) validates three essential characteristics of the numerical
module: (i)\,consistency between batch \texttt{\_ALL} subroutines and
their single-mode counterparts, (ii)\,satisfaction of mathematical
identities that hold analytically but are non-trivial to preserve
numerically, and (iii)\,agreement between the VSH inner-product
expressions and independent analytic forms derived from the
Geppert-Wiebicke coefficients. All tests were generated with suitable $\ell_\mathrm{max}$ determined by stability studies. Tests iterate over 15 uniformly
spaced interior co-latitude values $\theta \in (0, \pi)$ and
$\phi = \pi/4$, using double-precision arithmetic. The unnormalized Legendre routines are numerically stable to $\ell_\mathrm{max}=200$, and all other normalized routines are stable to $\ell_\mathrm{max}=2000$. 
Errors are reported as the maximum observed absolute difference over
all modes $(\ell, m)$ and all grid points.

%-----------------------------------------------------------------------
\subsubsection{Batch consistency}

%## UNNORMALIZED
\begin{figure}[h]
\centering
\includegraphics[width=0.45\textwidth]{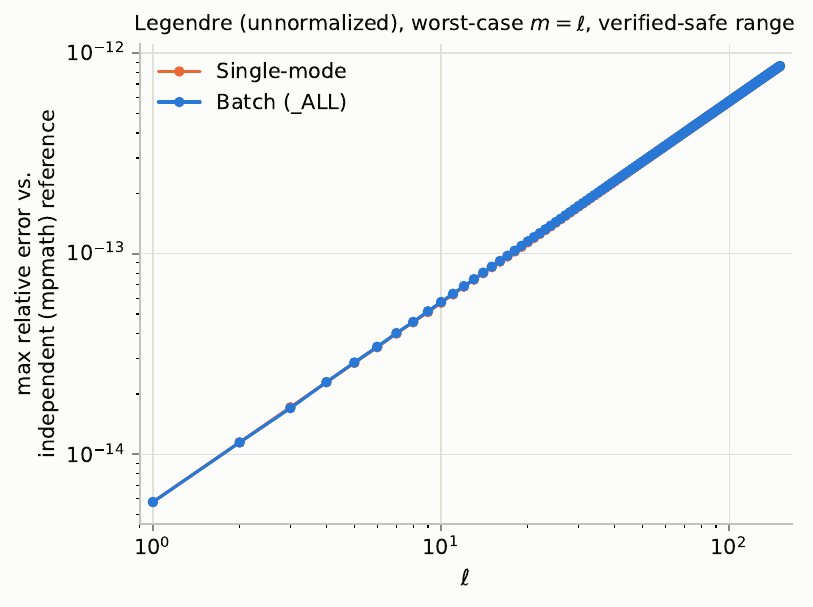}
\caption{Maximum relative error measured separately for batch subroutines in blue  (\texttt{ASSOC\_LEGENDRE\_ALL})
 and single-mode functions in orange (\texttt{ASSOC\_LEGENDRE}) for unnormalized Legendre function computations $P_{\ell}^{\ell}$, taken at the diagonal $m=\ell$ where cancellation error is the largest. Numerical error for single-mode and batch calculations determined through comparison to $P_{\ell}^m$ reference values computed in the Python \texttt{mpmath} library.}
\label{f:batchconsistency_unnorm}
\end{figure}

\begin{figure}[h]
\centering
\includegraphics[width=0.98\textwidth]{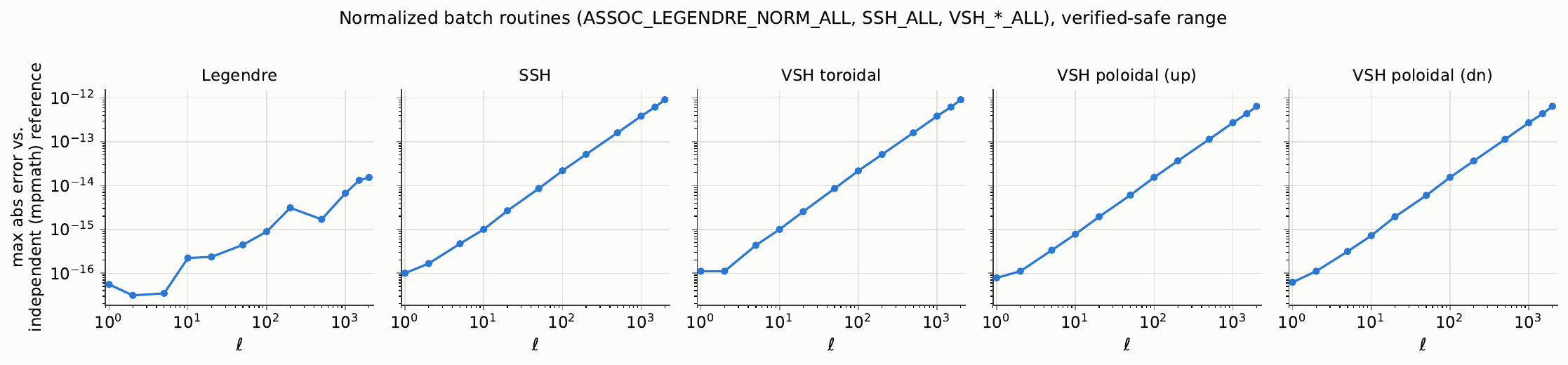}
\caption{Maximum absolute difference between batch \texttt{\_ALL} subroutines
  and corresponding reference values computed in the Python \texttt{mpmath} library; (left to right) \texttt{ASSOC\_LEGENDRE\_NORM}, \texttt{VSH\_TOR}, \texttt{VSH\_POL\_UP}, \texttt{VSH\_POL\_DN}.  Errors at or below $\sim\!10^{-15}$
  reflect double-precision rounding.}
\label{f:batchconsistency_norm}
\end{figure}

Each batch subroutine (suffixed with \texttt{\_ALL}) is compared pointwise against its
corresponding single-mode function.  For unnormalized Legendre routines, the
reported error is the absolute difference, and results are presented in Figure\,\ref{f:batchconsistency_unnorm}; for vector-valued VSH
routines the calculated error is the maximum absolute difference over the three spherical components
$(\hat{r}, \hat{\theta}, \hat{\phi})$, with results presented in Figure\,\ref{f:batchconsistency_norm}.

%## NORMALIZED

\begin{figure}[h]
\centering
\includegraphics[width=0.98\textwidth]{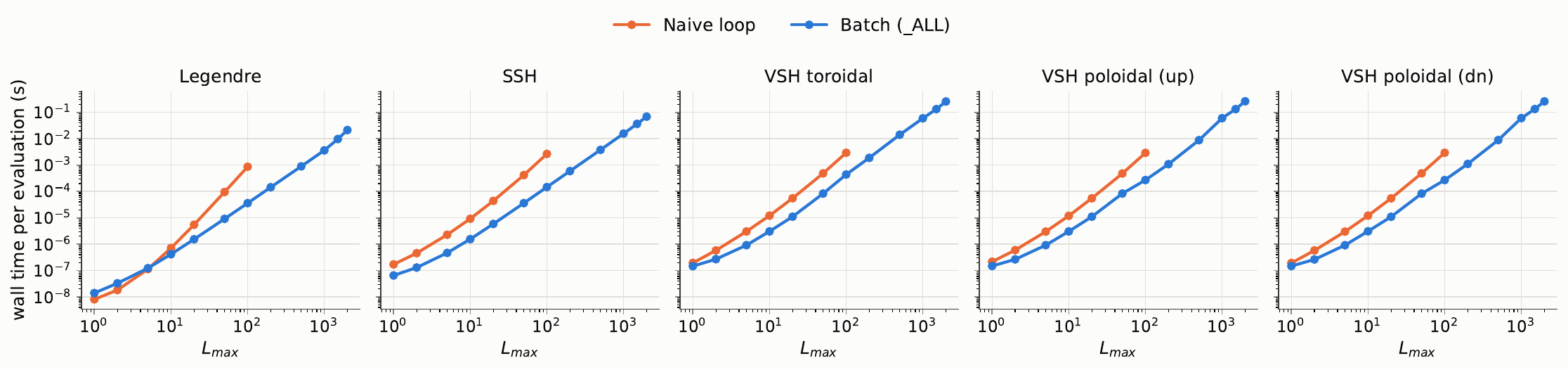}
\caption{Timing benchmarks for five fundamental FORTVSH routes, illustrating order-of-magnitude speedup of batch routines compared to single-mode routines.}
\label{f:benchmarkscaling}
\end{figure}

Figure\,\ref{f:benchmarkscaling} provides the timing benchmarks between batch subroutines and corresponding single-mode functions for the five numerical methods from Figure\,\ref{f:batchconsistency_norm}. Batch calculations outperform naive loops over single-mode computations by an order of magnitude in the overlapping range of stable $\ell$, and furthermore the batch routines extend the stable parameter range an additional order of magnitude to $\ell_\mathrm{max} = 2000$.

%-----------------------------------------------------------------------
\subsubsection{Mathematical identity tests: SSH orthonormality}

The orthonormality of the scalar spherical harmonics,
\begin{equation}
  \int Y_\ell^m(\theta,\phi)\,Y_{\ell'}^{m'*}(\theta,\phi)
  \,d\Omega = \delta_{\ell\ell'}\,\delta_{mm'},
\end{equation}
was verified numerically by evaluating the $\phi$ integral analytically
and approximating the $\cos\theta$ integral using the midpoint rule
with $N = 10{,}000$ quadrature points.
The maximum diagonal residual
$\bigl|\langle Y_{\ell m}, Y_{\ell m}\rangle - 1\bigr|$
is $9.1\times10^{-7}$, and the maximum off-diagonal residual
$\bigl|\langle Y_{\ell m}, Y_{\ell' m'}\rangle\bigr|$
for $(\ell,m) \neq (\ell',m')$ is $5.6\times10^{-7}$.
Both are consistent with the $\mathcal{O}(N^{-2})$ truncation error
of the midpoint rule and reflect quadrature precision, not
floating-point arithmetic errors.

\subsubsection{Mathematical identity tests: Poloidal-toroidal orthogonality.}
The bilinear dot product
$\mathbf{Y}_{J M}^{(+1)} \cdot \mathbf{Y}_{J M}^{(0)}$
vanishes analytically at every $(J, M, \theta, \phi)$ because
$\mathbf{Y}^{(0)}_{J M} \propto \hat{r} \times
\nabla_\omega Y_J^M$ is the image of
$\mathbf{Y}^{(+1)}_{J M} \propto \nabla_\omega Y_J^M$
under a $90^\circ$ rotation in the tangent plane.
The maximum observed value is $0$, confirming
that the batch routines reproduce this property to machine precision.

\subsubsection{Mathematical identity tests: VSH rotation inversion.}
The standard VSH $\mathbf{Y}_{J M}^{L}$ with $L = J \pm 1$ are
related to the polar VSH by the Clebsch-Gordan rotation
\begin{align}
  \mathbf{Y}_{J M}^{J+1} &=
    \sqrt{\tfrac{J}{2J+1}}\,\mathbf{Y}_{J M}^{(+1)}
    - \sqrt{\tfrac{J+1}{2J+1}}\,\mathbf{Y}_{J M}^{(-1)},
  \\[4pt]
  \mathbf{Y}_{J M}^{J-1} &=
    \sqrt{\tfrac{J+1}{2J+1}}\,\mathbf{Y}_{J M}^{(+1)}
    + \sqrt{\tfrac{J}{2J+1}}\,\mathbf{Y}_{J M}^{(-1)}.
\end{align}
The orthogonality of this rotation is verified by inverting it and
checking that \texttt{VSH\_POL\_UP\_ALL} and \texttt{VSH\_POL\_DN\_ALL}
reconstruct \texttt{PVSH\_POL\_ALL} and \texttt{PVSH\_RAD\_ALL}
exactly.  The maximum reconstruction residual for
$\mathbf{Y}_{J M}^{(+1)}$ is $1.7\times10^{-16}$, and for
$\mathbf{Y}_{J M}^{(-1)}$ is $1.1\times10^{-16}$.

%-----------------------------------------------------------------------
\subsubsection{Cross-validation against Geppert-Wiebicke forms}

Two VSH inner-product expressions that arise in the Hall-MHD
coupling integrals are shown in Eq.\,(\ref{e:GWIcoeff}) and Eq.\,(\ref{e:GWJcoeff}). These expressions are evaluated by two independent numerical routes and compared against
a closed-form analytic result derived from the Geppert-Wiebicke
coefficients.

\begin{figure}[h]
\centering
\includegraphics[width=0.98\textwidth]{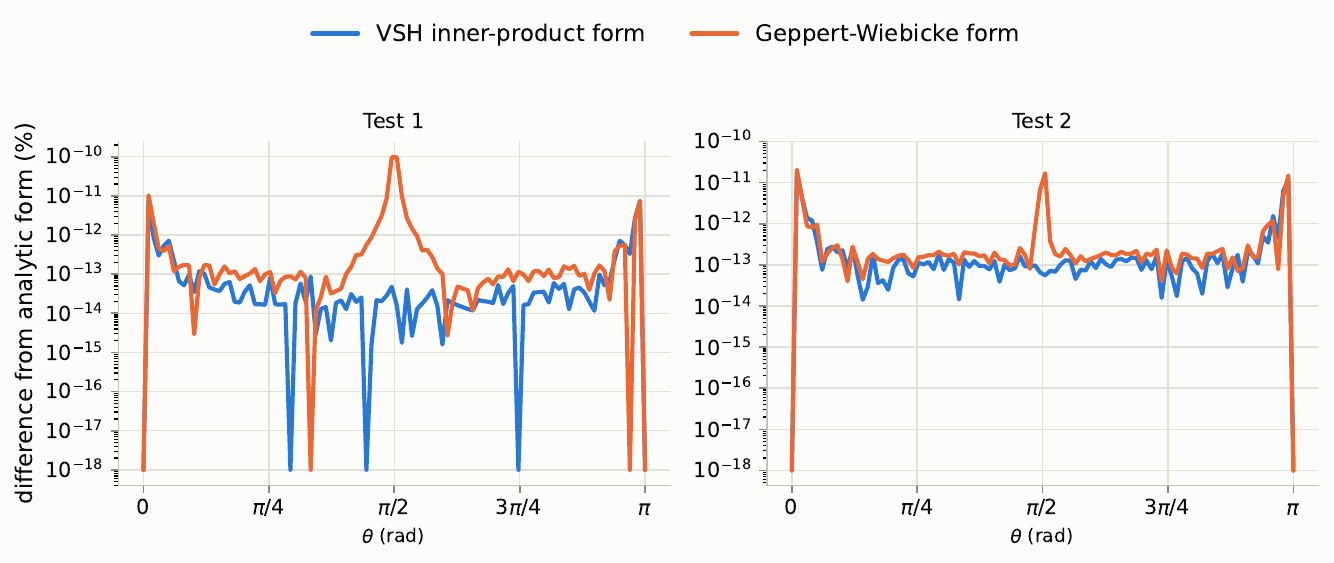}
\caption{Maximum relative deviation of the FORTVSH inner product calculation (blue) and the FORTVSH calculation of Geppert-Wiebicke coefficients (orange) cross-validated against closed-form analytic results. Left panel: $f_1\left(\theta,\phi\right)$ from Eq.\,(\ref{e:gwf1}). Right panel: $f_2\left(\theta,\phi\right)$ from Eq.\,(\ref{e:gwf2}).}
\label{f:crossval_accuracy}
\end{figure}

\noindent In Test 1 the product
$\left(\mathbf{Y}_{2,0}^2 \cdot \mathbf{Y}_{1,1}^1\right)
\left(\mathbf{Y}_{1,-1}^0 \cdot \mathbf{Y}_{0,0}^{1*}\right)$,
evaluated at $\phi = 0.123$\,rad, has the closed form
\begin{equation}\label{e:gwf1}
  f_1(\theta) =
  \frac{3\sqrt{5}}{64\pi^2}\sin^2\theta\cos^2\theta.
\end{equation}

\noindent In Test 2 the product
$\left(\mathbf{Y}_{2,0}^{(+1)} \cdot \mathbf{Y}_{3,2}^{(0)}\right)
\left(\mathbf{Y}_{1,0}^{(-1)} \cdot \mathbf{Y}_{2,-2}^{(-1*)}\right)$,
evaluated at $\phi = 1.006$\,rad, has the closed form
\begin{equation}\label{e:gwf2}
  f_2(\theta) =
  \frac{15\sqrt{105}}{128\pi^2}\,
  \frac{\sin^4\theta\cos^3\theta}{\sqrt{2}}\,e^{4i\phi}.
\end{equation}

Figure\,\ref{f:crossval_accuracy} displays the maximum relative deviation (\%) of the VSH route from the analytic value and the maximum relative deviation of the Geppert-Wiebicke route from the analytic value, as calculated across $\theta$. In both tests the VSH and Geppert-Wiebicke routes agree with the
analytic result to within a few units in the last place of double
precision, confirming that the two independently derived numerical
representations of the coupling integrals are mutually consistent and analytically precise.

\begin{figure}[h]
\centering
\includegraphics[width=0.45\textwidth]{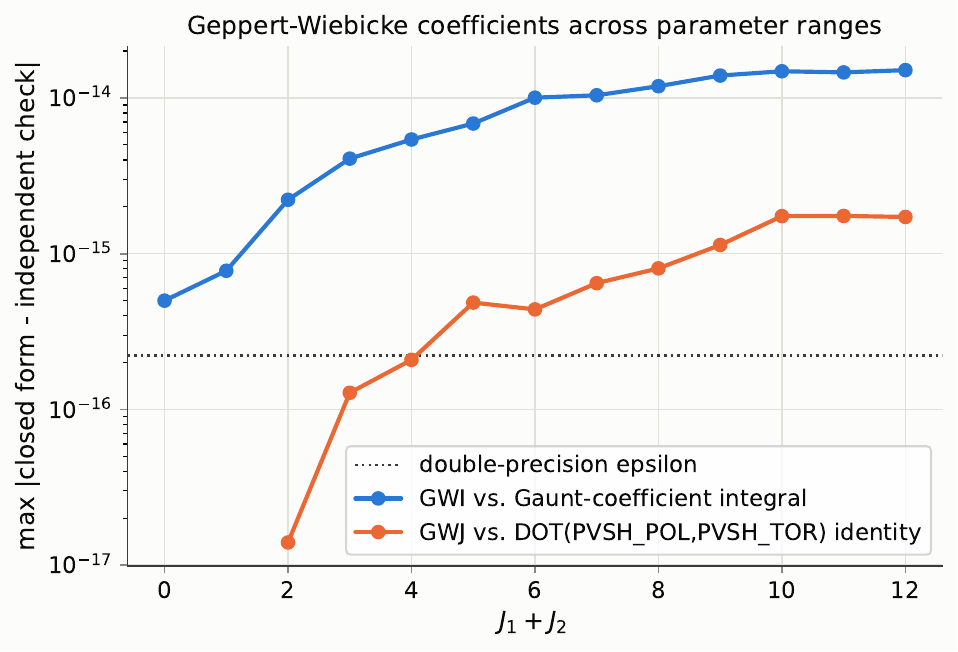}
\caption{Maximum absolute error per $J_1+J_2$ combination based on selection rules and triangle inequality. Dotted line at $2.22\times 10^{16}$ indicates machine epsilon for double precision. The \texttt{GWI} function outputs are cross-validated against independent quadrature-based integration results (blue), and the \texttt{GWJ} function outputs are compared to the analytically identical \texttt{PVSH\_POL(J2,M2)} and \texttt{PVSH\_TOR(J1,M1)} summed over all $L$ (orange).}
\label{f:gwsweep}
\end{figure}

Figure\,\ref{f:gwsweep} provides an aggregate view of numerical stability and precision through the parameter space up to $J_\mathrm{MAX}$=6. The FORTVSH function outputs from \texttt{GWI} and \texttt{GWJ} are compared to independent but analytically-identical forms on a grid of $N_{\theta},N_{\phi}=16,21$. Maximum absolute error is observed at the level of machine precision.

\subsection{Applications}\label{s:applications}

This section builds upon the fundamental validation and benchmarking in Section\,\ref{s:validation} by developing example use cases of incrementally increasing complexity. The examples provide evidence of the usefulness of FORTVSH for modeling spherical systems with a spectral approach.   

\subsubsection{Pure dipole synthesis}\label{s:application1}

\begin{figure}[ht]
\centering
\includegraphics[width=0.98\textwidth]{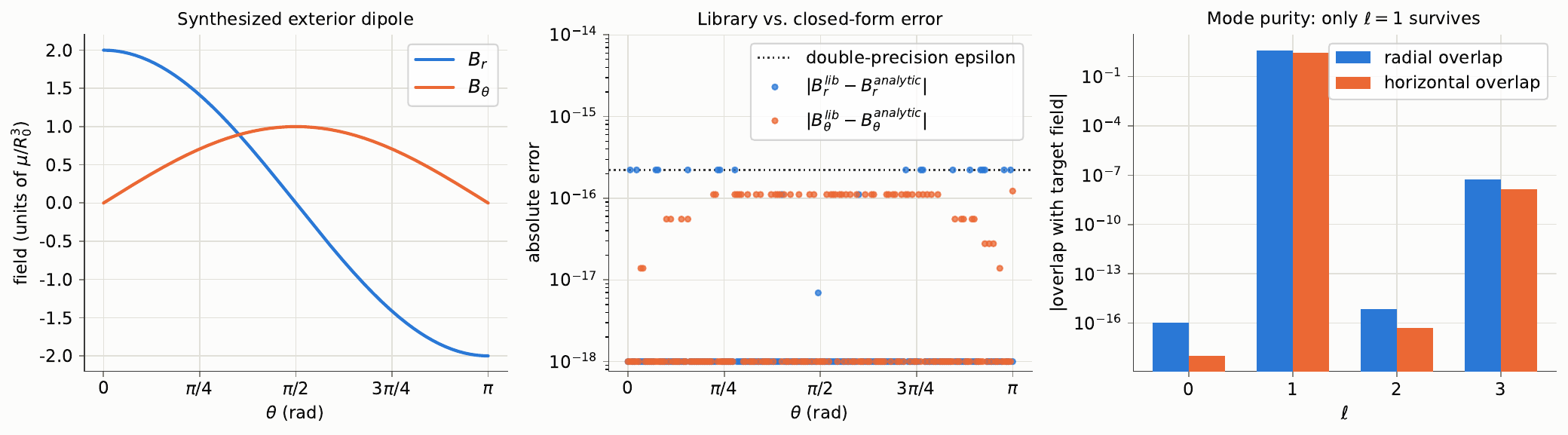}
\caption{Left: Poloidal magnetic dipole components $B_r\left(\theta\right)$ and $B_{\theta}\left(\theta\right)$ built in FORTVSH for the pure dipole synthesis analysis. Middle: Numerical error for $B_r\left(\theta\right)$ and $B_{\theta}\left(\theta\right)$ compared to analytic forms. Right: Illustration of spectral purity via calculation of component overlap with other $\ell$ modes.}
\label{f:dipoleaccuracy}
\end{figure}

A purely poloidal dipole field in vacuum is generated by invoking \texttt{PVSH\_RAD(1,0)} and \texttt{PVSH\_POL(1,0)} for the $B_r$ and $B_{\theta}$ components, respectively, inside the sphere. The interior B-field radial dependence and normalization are determined based on the scalar potential formulation such that the analytic forms
\begin{align}
B_r = & \frac{2\mu}{r^3}\cos\theta \\
B_{\theta} = & \frac{\mu}{r^3}\sin\theta
\end{align}

\noindent should perfectly match the numeric forms 
\begin{align}
B_r = & \frac{l\left(l+1\right)}{r^2}S\left(r\right)\times\texttt{PVSH\_RAD(l,m)}_r\label{e:brdipolefn} \\
B_{\theta} = & \frac{\sqrt{l\left(l+1\right)}}{r}\frac{dS}{dr}\times\texttt{PVSH\_POL(l,m)}_{\theta}\label{e:bthdipolefn}
\end{align}

\noindent if the dipole field is faithfully constructed, where $\mu=1$ and $S\left(r\right)=\sqrt{4\pi}/r\sqrt{3}$. Figure\,\ref{f:dipoleaccuracy} presents the angular distribution of the field components, the numerical error of the field construction routines, and evidence of mode purity. The FORTVSH basis functions produce the correct dipole solution at the level of machine precision and the solution has negligible spectral contamination in $\ell$.

\subsubsection{Magnetostatic boundary matching}\label{s:application2}

\begin{figure}[h]
\centering
\includegraphics[width=0.9\textwidth]{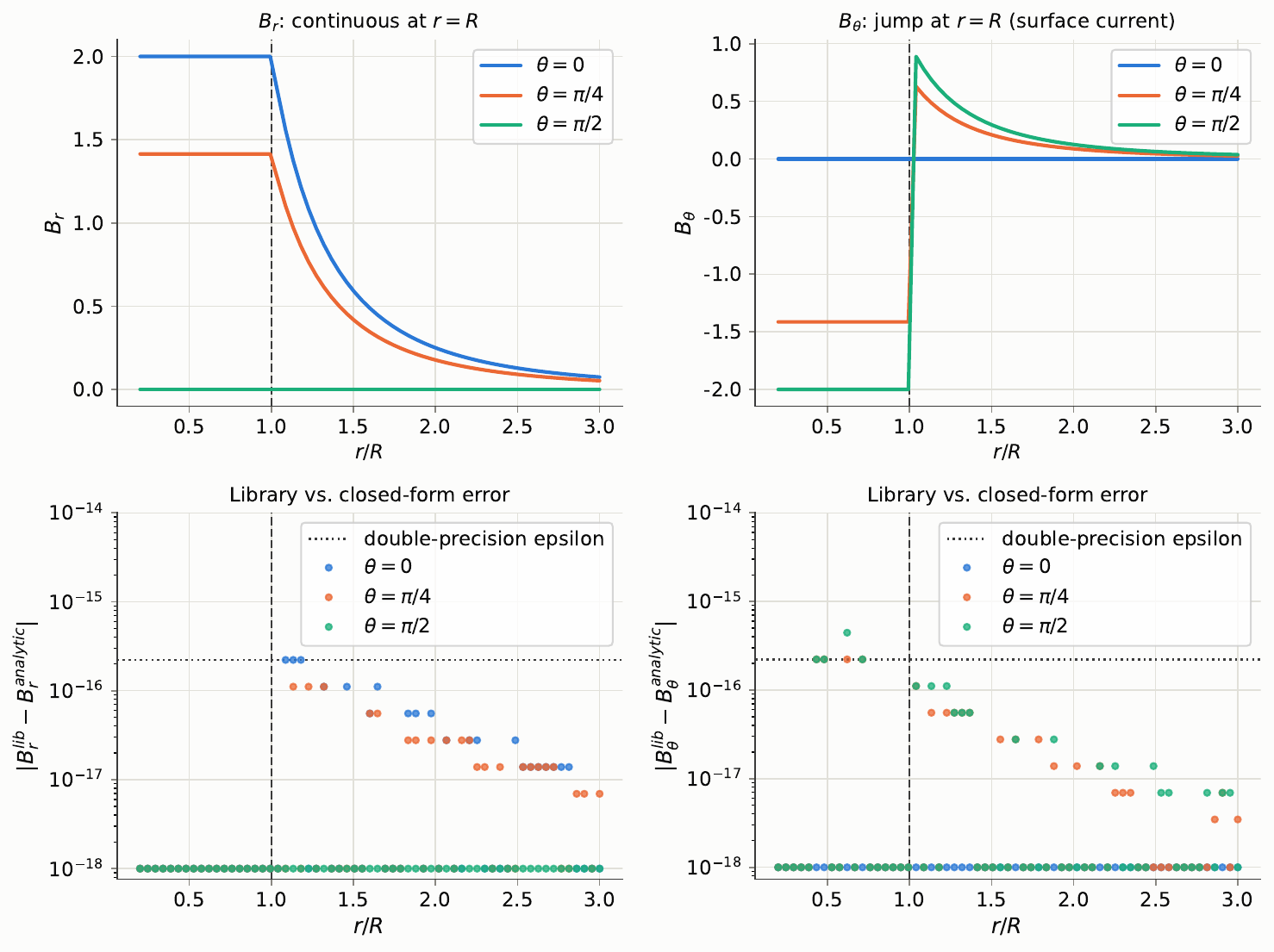}
\caption{Top row: $B_r(r)$ (left) and $B_{\theta}(r)$ (right) plotted in the sphere interior and exterior, highlight a continuous $B_r$ and discontinuous $B_{\theta}$ for the magnetostatic boundary matching analysis. Bottom row: absolute pointwise numerical error in $B_r(r)$ (left) and $B_{\theta}(r)$ compared to the analytic solution, plotted in the sphere interior and exterior.}
\label{f:dipolematch}
\end{figure}

The exterior dipole field from the previous example is now matched to a uniformly magnetized sphere at the boundary $r=R$. The interior field components have the functional structure as Eqs.\,(\ref{e:brdipolefn})-(\ref{e:bthdipolefn}), although $S(r)=Ar^2$. Figure\,\ref{f:dipolematch} shows that FORTVSH produces a continuous $B_r(R)$ along with the correct jump condition in $B_{\theta}$, accurately solving the boundary matching problem at or below the machine precision level.

\subsubsection{Spectral decomposition}\label{s:application4}

This example highlights FORTVSH capabilities performing spectral decomposition of arbitrary non-axisymmetric vector fields in a spherical coordinate system. The test field $\vec{B}_\mathrm{test}$ is composed of eight purely toroidal modes $\vec{B}_\mathrm{test}^{\ell m}$ which include all available $m$ for $\ell=1,2$ with each assigned equal weighting:
\begin{equation}
\vec{B}_\mathrm{test}^{\ell m} = A_{\ell m}^{\mathrm{tor}}\times\texttt{PVSH\_TOR(l,m)}
\end{equation}

\noindent with all $A_{\ell m}^{\mathrm{tor}} = 1$ by definition and where the reduced, component-wise algebraic expressions are therefore
\begin{align}
B_\mathrm{test}^{r} = & 0 \\
B_\mathrm{test}^{\theta} = & +i\sum_{\ell,m} \frac{A_{\ell m}^{\mathrm{tor}}}{\sqrt{\ell\left(\ell+1\right)}}\frac{1}{\sin\theta}\partial_{\phi}Y_{\ell}^m\\
B_\mathrm{test}^{\phi} = & -i\sum_{\ell,m} \frac{A_{\ell m}^{\mathrm{tor}}}{\sqrt{\ell\left(\ell+1\right)}}\partial_{\theta}Y_{\ell}^m
\end{align}
\begin{figure}[ht]
\centering
\includegraphics[width=0.98\textwidth]{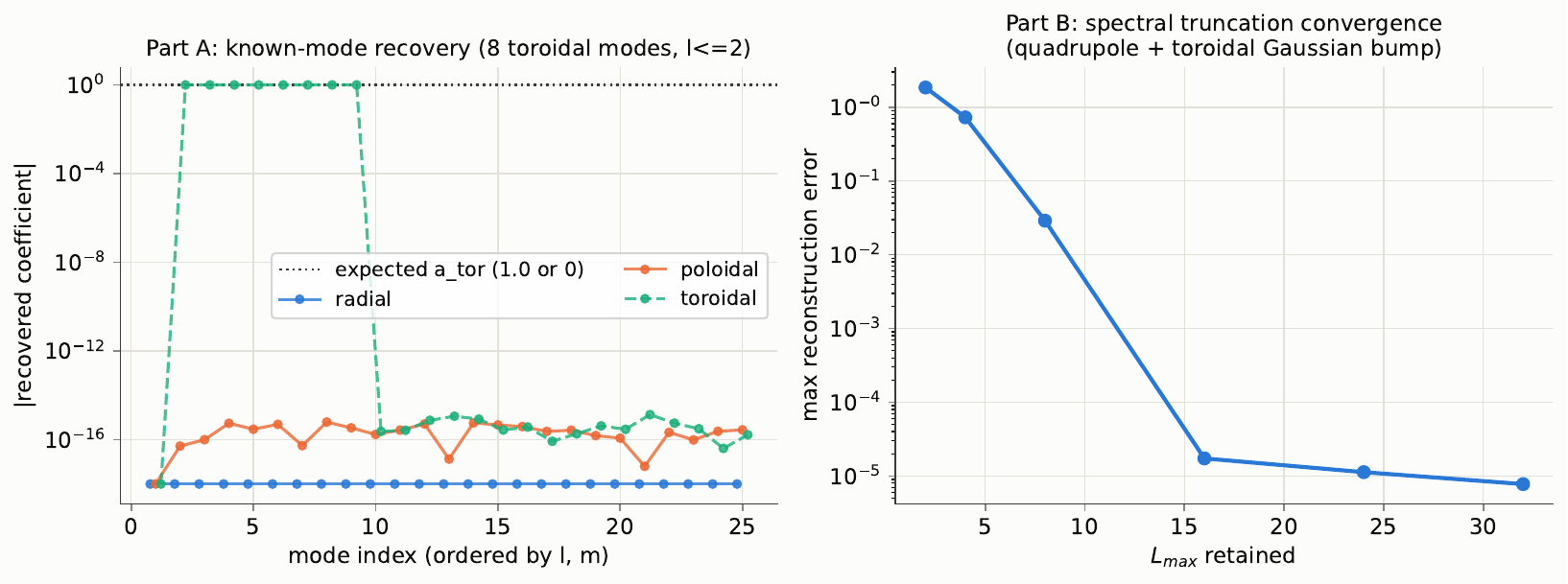}
\caption{Left: Results from Part A of the spectral decomposition analysis, showing precise calculations of supplied toroidal eigenmode amplitudes and no spectral contamination into other modes. Right: Spectral convergence results from Part B of the spectral decomposition analysis, showing characteristic knee near $\ell_\mathrm{max}=16$, consistent with evaluation grid configuration.}
\label{f:vsh_decomp}
\end{figure}

In Part A of this demonstration the FORTVSH routine \texttt{SHGLQ} is used to numerically integrate, using exact Gauss-Legendre polynomial quadrature in $\theta$, spectral coefficient expressions from Eqs.\,(\ref{e:vshcoeff1}) and (\ref{e:vshcoeff2}) to compute each $A_{\ell m}^{\mathrm{tor}}$. The full 2D quadrature leverages the \texttt{SHGLQ} routine alongside an \textit{ad hoc} uniform trapezoidal integration in $\phi$. The left panel of Figure\,\ref{f:vsh_decomp} shows the results. All original weights are perfectly recovered, and all non-contributing components have coefficients at the numerical noise floor. 

\begin{figure}[h]
\centering
\includegraphics[width=0.45\textwidth]{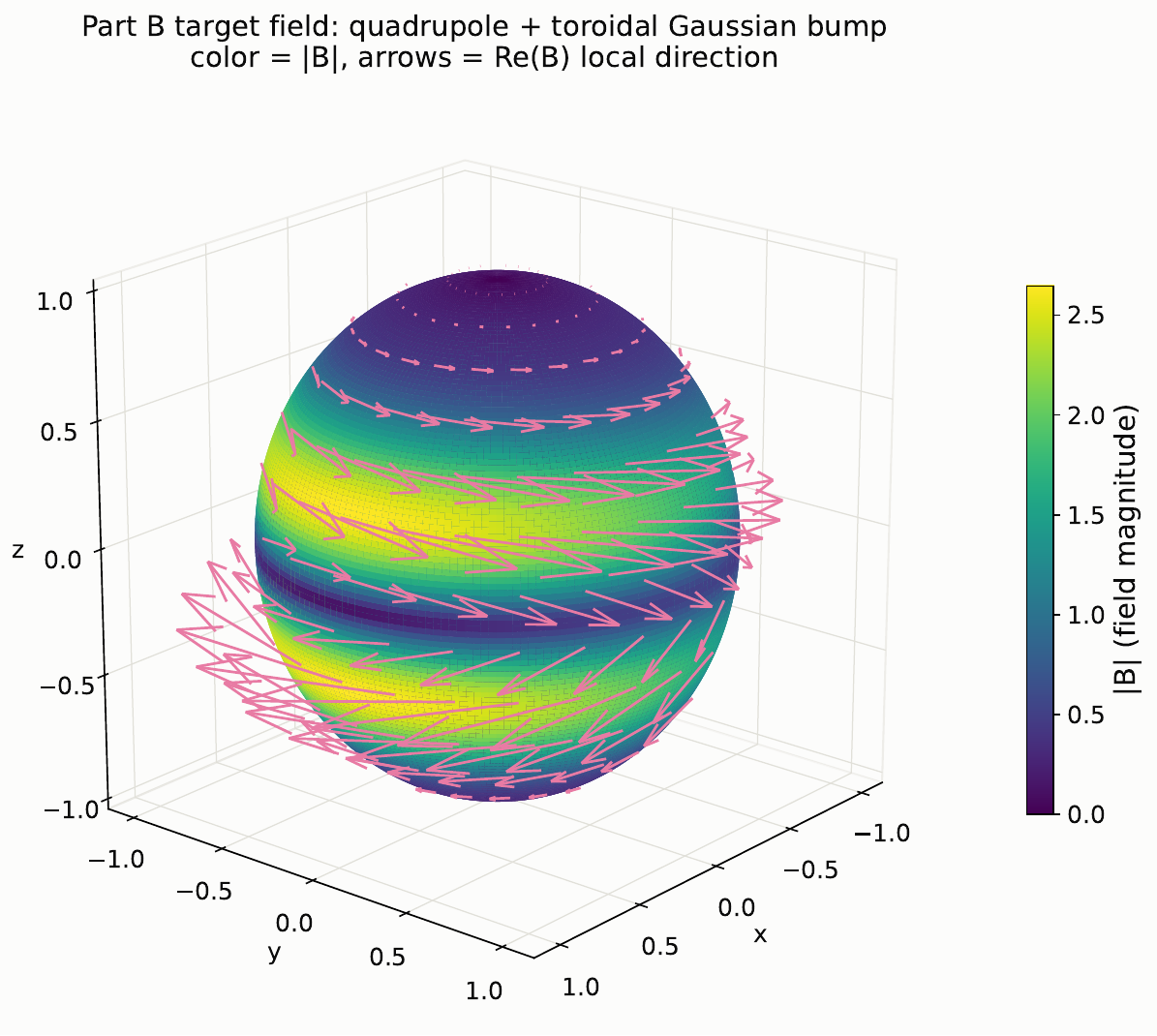}
\caption{3D representation of the imposed toroidal magnetic field in Part B of the spectral decomposition analysis. Color scale corresponds to field magnitude, and pink arrows indicate local direction of the real magnetic field component. Quadrupolar structure is clearly identifiable.}
\label{f:vsh_tor_3d}
\end{figure}

\begin{figure}[h]
\centering
\includegraphics[width=0.90\textwidth]{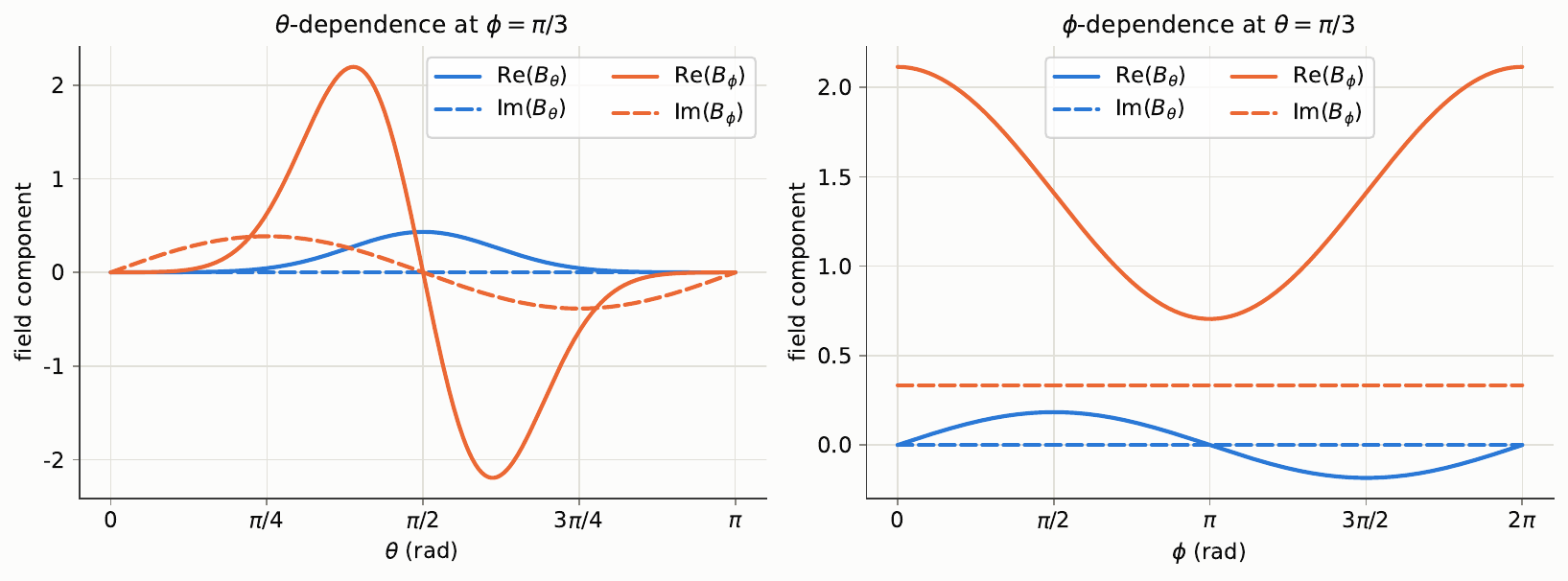}
\caption{2D slices of both complex components of the total toroidal magnetic field from Part B of the spectral decomposition analysis, including the background quadrupole structure and the Gaussian bump. All angular field components vanish at the poles.}
\label{f:vsh_tor_profiles}
\end{figure}

In Part B of this demonstration, the test's spectral bandwidth is increased as the discretized mode sum is replaced with a mode continuum. A purely toroidal quadrupole background field ($l=2, m=0$) is superimposed with a toroidal Gaussian bump with the analytic form
\begin{align}
B_\mathrm{bump}^{\theta} = & \frac{1}{2}\sin\theta\sin\phi\,e^{-\left(\theta-\pi/2\right)^2/2\sigma^2}\label{e:bthbump} \\
B_\mathrm{bump}^{\phi} = & \left(1+\frac{1}{2}\cos\phi\right)\left[\sin\left(2\theta\right)-\frac{\theta-\pi/2}{\sigma^2}\sin^2\theta \right]\,e^{-\left(\theta-\pi/2\right)^2/2\sigma^2}\label{e:bphbump}
\end{align}
which requires an infinite decomposition in $\ell$ and $m$ modes to perfectly reconstruct. Figure\,\ref{f:vsh_tor_3d} gives a high-level view of the field magnitude and vector direction, with a clear $B_{\phi}$ sign change (pink arrows) across the equator, consistent with a toroidal quadrupole. Figure\,\ref{f:vsh_tor_profiles} presents field component profiles at intermediate values of $\theta$ and $\phi$ where Eqs.\,(\ref{e:bthbump}) and (\ref{e:bphbump}) are nonzero. Both complex components are presented to illustrate smoothness and completeness.

The right panel of Figure\,\ref{f:vsh_decomp} shows the reconstruction error as a function of the permitted $\ell_\mathrm{max}$, which should vanish as $\ell_\mathrm{max}\rightarrow\infty$.

\section{Discussion}\label{s:discussion}

The summary of identities and relationships presented in this manuscript accompanies an installable, free, open-source Fortran package - FORTVSH - that provides performant, 
precise computations of VSH components within multiple standard representations. FORTVSH is designed for general applicability to problems in geophysics and astrophysics, but is especially suitable for vector field analysis under poloidal-toroidal decomposition. In the forward-transform context, where arbitrary continuous functions are decomposed into VSH modes, the VSH form the foundation of spectral 
numerical models designed to simulate 3D spherical dynamic systems, especially fluids and plasmas. This category of computer simulation is relevant for the internal magnetic field of neutron stars, astrophysical or planetary magnetospheres, dynamos, and laboratory plasma sources. In the inverse-transform context, VSH are integral for reconstituting spectra from discrete sampling (from computer simulations or \textit{in situ} instrumentation). The FORTVSH v1.0.0 library release contains numerical methods suitable only for forward-transform operations, although inverse-transform coverage is anticipated for a future release. Additional scheduled enhancements and features include support for Wigner $D$-matrices to enable simulation of multi-axis physics \textit{i.e.,} independent rotational and magnetic axes, implementation of C bindings via a \texttt{bind(c)} interface layer wrapping the FORTVSH array-valued functions, as well as Python bindings layered via \texttt{ctypes}, \texttt{cffi}, or \texttt{f2py}. 

The validation results presented in Section\,\ref{s:validation} demonstrate that FORTVSH computes VSH modes in the implemented representations to within machine precision. Batch routines agree with their single-mode counterparts and with independent double-precision reference values generated from the Python \texttt{mpmath} library, typically below $10^{-15}$ absolute error. Section\,\ref{s:validation} also demonstrates numerical preservation of essential analytic identities, including poloidal-toroidal orthogonality, VSH rotation inversion between the polar and $L^2$-eigenfunction bases, and cross-validation of the Geppert-Wiebicke coupling coefficients against independently-derived closed forms. The Legendre function recurrence is numerically stable to $\ell_\mathrm{max}=200$ in the unnormalized form and to $\ell_\mathrm{max}=2000$ using the normalized \cite{Holmes2002AUA} recurrence. Importantly this stability in $\ell$ is directly inherited by every SSH and VSH routine that is built upon it. Batch routines additionally outperform naive single-mode loops by an order of magnitude across the overlapping range of stable $\ell$, and extend the usable range a further order of magnitude beyond what single-mode evaluation can achieve. These results establish FORTVSH as a numerically reliable foundation for the spectral representation of vector fields in parameter ranges relevant to the geophysical and astrophysical systems motivating this work.

Several existing Fortran packages provide spherical harmonic infrastructure for geoscientific applications, but for two distinct structural reasons do not support poloidal-toroidal magnetic field representation. SHTOOLS \citep{SHTOOLS} offers a mature, optimized Fortran 95/Python archive supporting standard geodetic normalization and localized spectral analysis via Slepian functions. However its vector field routines are built around the gradient of a scalar potential, which is the appropriate representation for a curl-free field such as gravity or a current-free magnetic field configuration. A curl-free field has no toroidal component by construction, since a toroidal field is intrinsically rotational and not derivable from a scalar gradient. SHTOOLS therefore cannot represent a toroidal field at all, independent of application. SPHEREPACK \citep{spherepack3} provides a two-part vector spherical harmonic transform consisting of a rotational, toroidal-like stream function and a divergent, poloidal-like velocity potential, but this decomposition is confined to a purely tangential 2D field on a single shell. SPHEREPACK has been designed for modeling and analysis of horizontal atmospheric wind, which has no radial component. Therefore SPHEREPACK cannot express the coupling between a vector field's radial and tangential structure encoded within truly poloidal fields. 

FORTVSH instead implements the full radial-poloidal-toroidal VSH triad together through formal curl identities that map directly onto the induction equation's differential structure. FORTVSH implements the self-consistent VSH approach as a lightweight, dependency-free Fortran library with no Python layer, no FFT backend, and no BLAS/LAPACK requirement. These key features make FORTVSH directly embeddable, with minimal translation overhead, within purpose-built time-advance schemes for radially-stratified MHD systems. 

%% The following commands are for the statements about the availability of data sets and/or software code corresponding to the manuscript.
%% It is strongly recommended to make use of these sections in case data sets and/or software code have been part of your research the article is based on.

\codeavailability{FORTVSH is available on GitHub at
\url{https://github.com/justinelfritz/FORTVSH}, and comprehensive documentation (FORD) is available at \url{https://justinelfritz.github.io/FORTVSH/}. The package version described in this manuscript, v1.0.0, is archived on Zenodo at \url{https://doi.org/10.5281/zenodo.21855680} \citep{fortvshv100} and is distributed under the BSD 3-Clause license. 

All figures in this manuscript can be regenerated from the same Fortran source used for testing and validation. From the repository root, configure and build with the opt-in figure-data targets enabled: \\ \texttt{cmake -B build -DVSH\_BUILD\_EXAMPLES=ON} \texttt{-DVSH\_BUILD\_BENCHMARK=ON -DVSH\_BUILD\_CONVERGENCE=ON \\ -DVSH\_BUILD\_STABILITY=ON \&\& cmake --build build}. \\ Then run \texttt{cd build \&\& ctest --output-on-failure} to execute the core validation suite (\texttt{vsh\_test}), which writes \texttt{TEST1.dat} and \texttt{TEST2.dat} to the \texttt{validation/} directory and is checked for pass or fail. Remaining figure data comes from the opt-in executables, which must be run manually from the repository root to produce output in the expected relative directories: \texttt{./build/vsh\_benchmark} (to the \texttt{benchmark/} directory), \texttt{./build/vsh\_convergence} (to the \texttt{convergence/} directory), \texttt{./build/vsh\_stability} and \texttt{./build/vsh\_unnorm\_stability} (to the \texttt{stability/} directory), and the six demonstration executables \texttt{dipole\_synthesis}, \texttt{dipole\_uniform\_sphere}, \texttt{dipole\_flux\_expulsion}, \texttt{vsh\_decomposition}, \texttt{gwi\_gwj\_sweep}, and \\ \texttt{vsh\_decomposition\_tor} (each to their respective subdirectories within \texttt{examples/}). With \texttt{py/requirements.txt} installed (numpy, matplotlib, mpmath), each figure is then produced by its matching \texttt{py/plot\_*.py} script run from the repository root (\textit{e.g.}, \texttt{python3 py/plotError.py}, \texttt{python3 py/plot\_stability.py}, \texttt{python3 py/plot\_vsh\_decomposition\_tor.py}), which reads the corresponding \texttt{.dat} files and writes matched PDF/PNG pairs into the \texttt{/tex/Copernicus-EGU/figures/} directory.} %% use this section when having only software code available

%\dataavailability{TEXT} %% use this section when having only data sets available

%\codedataavailability{TEXT} %% use this section when having data sets and software code available

%\sampleavailability{TEXT} %% use this section when having geoscientific samples available

%\videosupplement{TEXT} %% use this section when having video supplements available

\appendix
\section{Numerical Module}\label{a:numericalmodule}
The Fortran library FORTVSH contains functions and subroutines that support all essential calculations detailed in this manuscript. Spatial arguments are
assumed to lie in the domains $-1 \leq X \leq +1$, $0 \leq \theta \leq \pi$,
$0 \leq \phi \leq 2\pi$.

This appendix describes the two tiers of evaluation that FORTVSH provides. \textit{Single-mode} functions
accept explicit $(\ell, m, \theta, \phi)$ arguments and return values for
a single mode evaluated at a specified location. \textit{Batch} subroutines (suffixed with \texttt{\_ALL}) accept an
\texttt{LMAX} argument and evaluate all modes
$0 \leq \ell \leq \ell_\mathrm{max}$, $-\ell \leq m \leq \ell$ at a single specified location
$(\theta, \phi)$. Batch subroutines store results in flat arrays indexed by
$\texttt{PLM\_INDEX}(\ell,m) = \ell(\ell{+}1)/2 + m + 1$ for associated
Legendre outputs or $\texttt{YLM\_INDEX}(\ell,m) = \ell^2 + \ell + m + 1$
for scalar spherical harmonic outputs; this indexing methodology is consistent with SHTOOLS usage \citep{SHTOOLS}. Vector-valued batch subroutines return arrays
of shape $(3, (\ell_\mathrm{max}{+}1)^2)$, where the array indices 1, 2, 3
correspond to $\hat{r}$, $\hat{\theta}$, $\hat{\phi}$, respectively.

\subsection{Legendre polynomials - single mode}
\begin{itemize}
    \item \texttt{LEGENDRE($\ell$,X)}: Evaluates the Legendre polynomial $P_\ell(X)$
    of order $\ell$ at $X = \cos\theta$ using an unnormalized approach via the Bonnet three-term recurrence relation $\left(\ell+1\right)P_{\ell+1}(X) = \left(2\ell+1\right) X P_{\ell}(X) - \ell P_{\ell-1}(X)$.
    \item \texttt{DDX\_LEGENDRE($\ell$,X)}: Evaluates the first derivative of the unnormalized Legendre polynomial $dP_\ell/dX$ at $X = \cos\theta$. Away from the poles the recurrence relation $\left(X^2-1\right)P_{\ell}'(X) = \ell\left[XP_{\ell}(X) - P_{\ell-1}(X)\right]$ is used, and at the poles $P_{\ell}'\left(X=\pm 1\right) = \frac{1}{2}\left(\pm1\right)^{\ell+1}\ell\left(\ell+1\right)$.
    \item \texttt{ASSOC\_LEGENDRE($\ell$,$m$,X)}: Evaluates the associated Legendre
    polynomial $P_\ell^m(X)$ of order $\ell$ and degree $m$ at $X = \cos\theta$, using the Condon-Shortley phase convention and Bonnet upward recurrence beginning at the diagonal term $P_{m}^m(X) = \left(-1\right)^m \left(2m-1\right)!!\left(1-X^2\right)^{m/2}$. For negative orders $-k$, $P_{\ell}^{-k}(X) = \left(-1\right)^k \frac{\left(\ell-k\right)!}{\left(\ell+k\right)!}P_{\ell}^k(X)$.
    \item \texttt{ASSOC\_LEGENDRE\_AND\_DERIV($\ell$,$m$,X,PLM,DPLM)}: Simultaneously returns
    both $P_\ell^m(X)$ and $dP_\ell^m/dX$ using a single, upward recurrence
    pass. This avoids redundant computations that would result from calling
    \texttt{ASSOC\_LEGENDRE} and \texttt{DDX\_ASSOC\_LEGENDRE} independently,
    and is the basis for subsequent gradient computations.
    \item \texttt{DDX\_ASSOC\_LEGENDRE($\ell$,$m$,X)}: Evaluates $dP_\ell^m/dX$ at
    $X = \cos\theta$. Implemented as a wrapper for \\
    \texttt{ASSOC\_LEGENDRE\_AND\_DERIV}.
\end{itemize}

\subsection{Legendre polynomials - batch}
\begin{itemize}
    \item \texttt{ASSOC\_LEGENDRE\_ALL(P,LMAX,X)}: Evaluates all unnormalized
    $P_\ell^m(X)$ for $0 \leq \ell \leq \ell_\mathrm{max}$, $0 \leq m \leq \ell$
    using the Bonnet three-term recurrence.
    \item \texttt{DDX\_ASSOC\_LEGENDRE\_ALL(DP,P,LMAX,X)}: Evaluates all
    $dP_\ell^m/dX$ given a precomputed array \texttt{P} from \\
    \texttt{ASSOC\_LEGENDRE\_ALL}.
    \item \texttt{ASSOC\_LEGENDRE\_NORM\_ALL(PNORM,LMAX,X)}: Evaluates all
    $4\pi$-normalized associated Legendre values
    $\bar{P}_\ell^m(X) = N_{\ell m}\,P_\ell^m(X)$, where
    $N_{\ell m} = \sqrt{(2\ell+1)(\ell-m)!\,/\,(4\pi\,(\ell+m)!)}$,
    using the \citet{Holmes2002AUA} modified forward-column recurrence.
    This algorithm incorporates $N_{\ell m}$ directly into the recurrence coefficients,
    keeping all intermediate values $\mathcal{O}(1/\sqrt{4\pi})$ and extending
    numerical stability to $\ell \approx 2000$ compared to $\ell \approx 200$ for
    the implementation of standard Bonnet recurrence.
    \item \texttt{DDX\_ASSOC\_LEGENDRE\_NORM\_ALL(DPNORM,PNORM,LMAX,X)}: Evaluates
    all $d\bar{P}_\ell^m/dX$ given precomputed \texttt{PNORM} from \texttt{ASSOC\_LEGENDRE\_NORM\_ALL}, using the relation
    \begin{equation*}
        \left(1-X^2\right)\frac{d\bar{P}_\ell^m}{dX} =
        \sqrt{\tfrac{(2\ell+1)(\ell^2-m^2)}{2\ell-1}}\;\bar{P}_{\ell-1}^m
        - \ell X\,\bar{P}_\ell^m.
    \end{equation*}
\end{itemize}

\subsection{Scalar spherical harmonics - single mode}
\begin{itemize}
    \item \texttt{SSH($\ell$,$m$,$\theta$,$\phi$)}: Evaluates $Y_\ell^m(\theta,\phi)$ using unnormalized single-mode associated Legendre polynomial $P_{\ell}^m$ such that $Y_\ell^m(\theta,\phi) = \sqrt{\frac{2\ell+1}{4\pi}\frac{\left(\ell-m\right)!}{\left(\ell+m\right)!}}P_\ell^m(\cos\theta)e^{i m \phi}$.
    \item \texttt{GRAD\_SSH($\ell$,$m$,$\theta$,$\phi$)}: Evaluates the angular
    gradient $\vec{\nabla}_{\omega} Y_\ell^m$ using the $Y_\ell^m$ form described in the \texttt{SSH} routine definition.
    \item \texttt{L\_SSH($\ell$,$m$,$\theta$,$\phi$)}: Evaluates
    $\hat{r}\times\vec{\nabla}_{\omega} Y_\ell^m$ using the $Y_\ell^m$ form described in the \texttt{SSH} routine definition.
\end{itemize}

\subsection{Scalar spherical harmonics - batch}
\begin{itemize}
    \item \texttt{SSH\_ALL(YLM,LMAX,$\theta$,$\phi$)}: Evaluates all
    $Y_\ell^m(\theta,\phi)$ with $m\ge 0$ using a single call to \texttt{ASSOC\_LEGENDRE\_NORM\_ALL}. Negative-$m$ entries are filled using the conjugate symmetry $Y_\ell^{-m} = (-1)^m\,Y_\ell^{m*}$.
    \item \texttt{GRAD\_SSH\_ALL(OUT,LMAX,$\theta$,$\phi$)}: Evaluates all
    $\vec{\nabla}_{\omega} Y_\ell^m$ using calls to \texttt{ASSOC\_LEGENDRE\_NORM\_ALL} and \linebreak \texttt{DDX\_ASSOC\_LEGENDRE\_NORM\_ALL}.
    \item \texttt{L\_SSH\_ALL(OUT,LMAX,$\theta$,$\phi$)}: Evaluates all
    $\hat{r}\times\vec{\nabla}_{\omega} Y_\ell^m$ using calls to \texttt{ASSOC\_LEGENDRE\_NORM\_ALL} and \linebreak \texttt{DDX\_ASSOC\_LEGENDRE\_NORM\_ALL}.
\end{itemize}

\subsection{Polar vector spherical harmonics - single mode}
\begin{itemize}
    \item \texttt{PVSH\_RAD(J,M,$\theta$,$\phi$)}: Evaluates $\mathbf{Y}_{J M}^{(-1)}$ using the \texttt{SSH} routine.
    \item \texttt{PVSH\_TOR(J,M,$\theta$,$\phi$)}: Evaluates $\mathbf{Y}_{J M}^{(0)}$ using the \texttt{ASSOC\_LEGENDRE\_AND\_DERIV} routine.
    \item \texttt{PVSH\_POL(J,M,$\theta$,$\phi$)}: Evaluates $\mathbf{Y}_{J M}^{(+1)}$ using the \texttt{ASSOC\_LEGENDRE\_AND\_DERIV} routine. 
\end{itemize}

\subsection{Standard vector spherical harmonics - single mode}
\begin{itemize}
    \item \texttt{VSH\_TOR(J,M,$\theta$,$\phi$)}: Evaluates $\mathbf{Y}_{J M}^{L}$.
    Identical to \texttt{PVSH\_TOR}.
    \item \texttt{VSH\_POL\_DN(J,M,$\theta$,$\phi$)}: Evaluates $\mathbf{Y}_{J M}^{L-1}$ using the \texttt{ASSOC\_LEGENDRE\_AND\_DERIV} routine.
    \item \texttt{VSH\_POL\_UP(J,M,$\theta$,$\phi$)}: Evaluates $\mathbf{Y}_{J M}^{L+1}$ using the \texttt{ASSOC\_LEGENDRE\_AND\_DERIV} routine.
\end{itemize}

\subsection{Vector spherical harmonics - batch}
All batch VSH subroutines share the output layout described above. Computations are achieved through calls to \linebreak\texttt{ASSOC\_LEGENDRE\_NORM\_ALL} and \texttt{DDX\_ASSOC\_LEGENDRE\_NORM\_ALL}. Toroidal and poloidal entries are zero for $\ell = 0$.
\begin{itemize}
    \item \texttt{PVSH\_RAD\_ALL(OUT,LMAX,$\theta$,$\phi$)}: Evaluates all
    $\mathbf{Y}_{J M}^{(-1)}$.
    \item \texttt{PVSH\_TOR\_ALL(OUT,LMAX,$\theta$,$\phi$)}: Evaluates all
    $\mathbf{Y}_{J M}^{(0)}$.
    \item \texttt{PVSH\_POL\_ALL(OUT,LMAX,$\theta$,$\phi$)}: Evaluates all
    $\mathbf{Y}_{J M}^{(+1)}$.
    \item \texttt{VSH\_TOR\_ALL(OUT,LMAX,$\theta$,$\phi$)}: Evaluates all
    $\mathbf{Y}_{J M}^{L}$.
    \item \texttt{VSH\_POL\_UP\_ALL(OUT,LMAX,$\theta$,$\phi$)}: Evaluates all
    $\mathbf{Y}_{J M}^{L+1}$.
    \item \texttt{VSH\_POL\_DN\_ALL(OUT,LMAX,$\theta$,$\phi$)}: Evaluates all
    $\mathbf{Y}_{J M}^{L-1}$.
\end{itemize}

\subsection{Angular momentum algebra}
\begin{itemize}
    \item \texttt{CGCOEFF(J1,M1,J2,M2,J3,M3)}: Computes the Clebsch-Gordan
    coefficient $C_{J_1 M_1 J_2 M_2}^{J_3 M_3}$.
    \item \texttt{SYMBOL3J(J1,M1,J2,M2,J3,M3)}: Computes the Wigner $3j$ symbol
    $\begin{pmatrix}J_1 & J_2 & J_3 \\ M_1 & M_2 & M_3\end{pmatrix}$.
    \item \texttt{GWI(J1,M1,J2,M2,L,M)}: Computes the $I_{J_1 M_1 J_2 M_2}^{L M}$
    coefficient (see Eq.\,\ref{e:GWIcoeff}).
    \item \texttt{GWJ(J1,M1,J2,M2,L,M)}: Computes the $J_{J_1 M_1 J_2 M_2}^{L M}$
    coefficient (see Eq.\,\ref{e:GWJcoeff}).
\end{itemize}

\subsection{Quadrature and utilities}
\begin{itemize}
    \item \texttt{SHGLQ(ZERO,W,LMAX)}: Returns $\ell_\mathrm{max}+1$ Gauss-Legendre
    quadrature nodes $x_i = \cos\theta_i \in [-1,1]$ (ordered from $-1$ to $+1$)
    and weights $w_i$ satisfying
    $\int_{-1}^{1} f(x)\,dx = \sum_i w_i\,f(x_i)$ exactly for all polynomials
    of degree $\leq 2\ell_\mathrm{max}+1$. Nodes are computed as zeros of
    $P_{\ell_\mathrm{max}+1}(x)$ via Newton iteration; weights via
    $w_i = 2\,/\!\left[(1-x_i^2)\left(P'_{\ell_\mathrm{max}+1}(x_i)\right)^2\right]$.
    \item \texttt{DOT(VSH1,VSH2)}: Evaluates the bilinear product
    $\sum_{i=1}^{3}\mathtt{VSH1}_i\,\mathtt{VSH2}_i$ of two complex 3-vectors.
    Note this is not the Hermitian inner product; use \texttt{DOT(A,CONJG(B))}
    where conjugation is required.
    \item \texttt{PLM\_INDEX($\ell$,$m$)}: Returns the 1D index $\ell(\ell{+}1)/2 + m + 1$
    for entry $(\ell, m)$ in any Legendre batch output array, valid for
    $0 \leq m \leq \ell$.
    \item \texttt{YLM\_INDEX($\ell$,$m$)}: Returns the 1D index $\ell^2 + \ell + m + 1$
    for entry $(\ell, m)$ in any spherical harmonic batch output array, valid for
    $-\ell \leq m \leq \ell$.
    \item \texttt{LOG\_FACT(N)}: Returns $\ln(N!)$ via $\ln\Gamma(N{+}1)$. Used
    throughout to avoid factorial overflow at large $\ell$.
\end{itemize}

%\subsection{}     %% Appendix A1, A2, etc.

\noappendix       %% use this to mark the end of the appendix section. Otherwise the figures might be numbered incorrectly (e.g. 10 instead of 1).

%% Regarding figures and tables in appendices, the following two options are possible depending on your general handling of figures and tables in the manuscript environment:

%% Option 1: If you sorted all figures and tables into the sections of the text, please also sort the appendix figures and appendix tables into the respective appendix sections.
%% They will be correctly named automatically.

%% Option 2: If you put all figures after the reference list, please insert appendix tables and figures after the normal tables and figures.
%% To rename them correctly to A1, A2, etc., please add the following commands in front of them:

\appendixfigures  %% needs to be added in front of appendix figures

\appendixtables   %% needs to be added in front of appendix tables

%% Please add \clearpage between each table and/or figure. Further guidelines on figures and tables can be found below.

\authorcontribution{The author is responsible for developing the physics context, mathematical models, and Fortran routines in the FORTVSH library, as well as the validation scheme and build infrastructure. The author is also responsible for maintenance of the FORTVSH library. The author reviewed and verified all artifacts from Claude Code, and takes full responsibility for all content in this manuscript and the accompanying software package.} %% this section is mandatory

\competinginterests{No competing interests are present.} %% this section is mandatory even if you declare that no competing interests are present

%\disclaimer{TEXT} %% optional section

\begin{acknowledgements}
The Anthropic Claude model (Claude Code) assisted the author with automation of numerical validation scripts and routines, standardizing package artifacts to align with build best practices, outlining drafts of manuscript text subsequently reviewed and revised by the author, and independent build verification and continuous-integration (CI) verification across compiler toolchains.
\end{acknowledgements}

%% REFERENCES
\bibliographystyle{copernicus}
\bibliography{FORTVSH.bib}

\end{document}